\documentclass[11pt,twoside]{article}
\usepackage{asp2010}

\resetcounters

\begin{document}

\title{New Organizations to Support Astroinformatics and Astrostatistics}
\author{Eric D. Feigelson$^1$, \v{Z}eljko Ivezi\'c$^2$, Joseph Hilbe$^3$, Kirk D. Borne$^4$
\affil{$^1$Dept. of Astronomy \& Astrostatistics and Center for Astrostatistics, Penn State University, 525 Davey Laboratory, University Park PA 16802}
\affil{$^2$Dept. of Astronomy, University of Washington, 3910 15th Ave NE, Seattle WA 98185 }
\affil{$^3$Dept. of Mathematical and Statistical Sciences, Arizona State University, P.O.Box 871804, Tempe, AZ 85287}
\affil{$^4$School of Physics, Astronomy \& Computational Sciences, George Mason University, 4400 University Drive, Fairfax VA 22030}}
\begin{abstract}
In the past two years, the environment within which astronomers conduct their data analysis and management has rapidly changed.  Working Groups associated with international societies and Big Data projects have emerged to support and stimulate the new fields of astroinformatics and astrostatistics.  Sponsoring societies include the Intenational Statistical Institute, International Astronomical Union, American Astronomical Society, and Large Synoptic Survey Telescope project.  They enthusiastically support cross-disciplinary activities where the advanced capabilities of computer science, statistics and related fields of applied mathematics are applied to advance research on planets, stars, galaxies and the Universe.  The ADASS community is encouraged to join these organizations and to explore and engage in their public communication Web site, the Astrostatistics and Astroinformatics Portal (http://asaip.psu.edu).  
\end{abstract}

\section{Introduction}

Astronomers are increasingly turning attention to advanced methodologies for data and science analysis, particularly for large-scale surveys. The use of sophisticated statistical and computational methods is the purview of the cross-disciplinary fields astroinformatics and astrostatistics developed by astronomers, computer scientists and statisticians. Examples of recent activity include use of wavelets and independent component analysis for interpreting the cosmic microwave background, Bayesian model selection for estimating the number of planets orbiting a star, clarified calculation of nondetections in photon counting instruments, data mining classification of irregularly spaced light curves, and adaption of various computations (radio interferometer correlations, mega-database management, astrophysical fluid and N-body simulations) to computers based on graphics processing units.  

We report here some changes in the environment within which astronomers conduct such activities.  A new refereed journal {\it Astronomy \& Computing} has been formed by the Elsevier publishing company.  New educational resources are emerging including texts oriented towards astronomy graduate students and researchers on statistical methodology (Wall \& Jenkins 2012, Feigelson \& Babu 2012a, Feigelson \& Babu 2012b) and data mining techniques (Way et al. 2011, Ivezi\'c et al. 2013).  The Virtual Observatory organization conducts outreach training, and week-long summer schools for student training in statistics and informatics are growing.   We concentrate on the recent emergence four organizations formed by societies and projects to advance astroinformatics and astrostatistics.  

\section{Four Organizations}

The astroinformatics and astrostatistics subcommunities have often operated as small semi-independent expert subcommunities in narrow areas such as large databases, data mining, algorithms, data visualization, semantic astronomy, instrument pipelines, telescope archives, data reduction, image processing, astrophysical model fitting, Poissonian data analysis, time series analysis, and several subspecialities of observational cosmology such as weak lensing and cosmological parameter estimation.  With the assistance of international societies with broad perspectives, these subcommunities can increasingly synergize with each other and with cross-disciplinary experts.   The recently created rganizations are:

\begin{description}

\item[International Statistical Institute (ISI)]  In 2009, the ISI (sister organization to the IAU) formed an {\it Astrostatistics Committee}, and in 2010 an {\it Astrostatistics Network}, under the leadership of Joseph Hilbe. It is now being reorganized as a permanent and independent {\it International Astrostatistics Association} that will serve as an Associated Society of the ISI. 

\item[International Astronomical Union (IAU)]  In August 2012, the IAU Commission 5 formed a {\it Working Group in Astrostatistics and Astroinformatics} under the leadership of Eric Feigelson.

\item[American Astronomical Society (AAS)]  In June 2012, the U.S. AAS Council formed a {\it Working Group in Astroinformatics and Astrostatistics} under the leadership of \v{Z}eljko Ivezi\'c.

\item[Large Synoptic Survey Telescope (LSST)] In 2010, the LSST project, recently recommended by the U.S. National Science Board to become the largest ground-based project supported by the National Science Foundation,   formed the {\it Informatics and Statistics Science Collaboration}  under the leadership of Kirk Borne to promote cross-disciplinary data science methodology for the project (Borne 2010).  

\end{description}

\section{Goals}

The above organizations have somewhat different orientations, and have yet to solidify their exact goals and activities.  But a general set of goals that reflects the interests and plans of these organizations would include the following:

\begin{description}
\item[TO PROMOTE] wide and knowledgeable use of known advanced statistical and  computational methods in astronomical research
\item[TO ENCOURAGE] development of new user-friendly algorithms, computational techniques, and statistical methods to address forefront astronomical research  problemsÊ
\item[TO FACILITATE] access to advanced software for the visualization, statistical  analysis, and computation involving astronomical data
\item[TO ORGANIZE] multi-disciplinary gatherings in astrostatistics and  astroinformatics
\item[TO ÊFOSTER] collaboration between astrophysicists, statisticians, and information scientists
\item[TO PROVIDE] educational and professional resources to the several scholarly  communities to further the above goals.
\end{description}

\section{Astrostatistics and Astroinformatics Portal}

The {\it Astrostatistics and Astroinformatics Portal} (ASAIP, http://asaip.psu.edu) is now emerging as the primary Web site to serve these organizations and the wider cross-disciplinary research communities.  ASAIP provides a large searchable collection of recent papers, discussion forums, informal articles, list of meetings, and other resources.  Over 250 papers are currently listed, with many new listings added each month.  The portal lists twenty pertinent meetings (conferences and workshops) that were held worldwide during 2012.  The online resources at ASAIP include recent books on informatics and statistics (including all those oriented specifically for astronomy), online courses (some with tuition and some at no cost), metasites with extensive resources (such as KDNuggets.com and .astronomy), blogs on statistics and informatics, links to research groups, astronomical test datasets, and job opportunities.  ASAIP currently has $\sim$300 members.  It is hosted by Penn State's Eberly College of Science with Eric Feigelson and Joseph Hilbe as editors.  Astronomers and related experts are encouraged to become members of ASAIP, giving them rights to contribute to forums, and to add articles, meetings and links.   

\section{Call for participation}

The ADASS community is strongly encouraged to join one or more of these organizations, and become members of ASAIP.  Their near-simultaneous emergence of represents a strong endorsement for the importance of sophisticated quantitative data analysis, both for ordinary and Big Data astronomical research problems, by leaders of major  societies and projects.    Cross-fertilization with expertise residing in allied fields such as applied mathematics, computer science and engineering, applied and mathematical statistics, geostatistics and geoinformatics has been limited in the past, and can be improved. The ISI, IAU, AAS, and LSST organizations represent a new phase in the development of our field, promoting greater communication, expertise, and collaboration on the difficult methodological challenges that face modern data-intensive astronomical research.

\bigskip

\acknowledgements We would like to thank the leaders of the International Statistical Institute, International Astronomical Union, American Astronomical Society, Large Synoptic Survey Telescope, and Penn State Eberly College of Science for their enthusiasm and energy in creating these organizations to further research and accomplishments in astroinformatics and astrostatistics. 


\bigskip

\noindent{\bf References}

Borne, K. D. 2010, {\it Astroinformatics: Data-Oriented Astronomy Research and Education}, Journal of Earth Science Informatics, 3, 5-17 \\

Feigelson, E. D. \& Babu, G. J. 2012a, {\it Modern Statistical Methods for Astronomy with R Applications}, Cambridge:Cambridge University Press \\

Feigelson, E. D. \& Babu, G. J. (eds.) 2012b, {\it Statistical Challenges in Modern Astronomy V}, New York:Springer \\

Ivezi\'c, \v{Z}., Connolly, A., VanderPlas, J. \& Gray, A. 2013, {\it Statistics, Data Mining, and Machine Learning in Astronomy:
      A Practical Python Guide for the Analysis of Survey Data}, Princeton:Princeton University Press \\

Wall, J. V. \& Jenkins, C. R. 2012, {\it Practical Statistics for Astronomers}, 2nd ed., Cambridge:Cambridge University Press \\

Way, M. J., Scargle, J. D., Ali, K. M. \& Srivastava, A. N. (eds.) {\it Advanced in Machine Learning and Data Mining for Astronomy}, London:Chapman \& Hall/CRC \\

\end{document}